\documentclass[prd,preprint,superscriptaddress,showpacs,byrevtex]{revtex4}
\usepackage{bm}
\def\fsl#1{\setbox0=\hbox{$#1$}           
   \dimen0=\wd0                                 
   \setbox1=\hbox{/} \dimen1=\wd1               
   \ifdim\dimen0>\dimen1                        
      \rlap{\hbox to \dimen0{\hfil/\hfil}}      
      #1                                        
   \else                                        
      \rlap{\hbox to \dimen1{\hfil$#1$\hfil}}   
      /                                         
   \fi}                                         %
\newcommand{\be}{\begin{equation}}
\newcommand{\ee}{\end{equation}}
\newcommand{\bea}{\begin{eqnarray}}
\newcommand{\eea}{\end{eqnarray}}
\newcommand{\beq}{\begin{equation}}
\newcommand{\eeq}{\end{equation}}
\newcommand{\beqs}{\begin{eqnarray}}
\newcommand{\eeqs}{\end{eqnarray}}

\begin{document}
\title{ Coulomb Potential Is Not a Part of The QCD Potential }
\author{Gouranga C Nayak }\thanks{G. C. Nayak was affiliated with C. N. Yang Institute for Theoretical Physics in 2004-2007.}
\affiliation{ C. N. Yang Institute for Theoretical Physics, Stony Brook University, Stony Brook NY, 11794-3840 USA}
\date{\today}
\begin{abstract}
The Coulomb plus linear potential is widely used in QCD. However, in this paper we show that the Coulomb potential of the form $\frac{1}{r}$ is not a part of the QCD potential. This is because the form $\frac{g^2}{r}$ is for abelian theory (not QCD) and the form $\frac{g^2(\mu)}{r}$ in QCD at short distance is not of the Coulomb form $\frac{1}{r}$ because  $g(\mu)$ depends on the mass/length scale $\mu$. Similarly at long distance the QCD potential corresponds to the potential in the classical Yang-Mills theory which does not have the Coulomb form $\frac{1}{r}$ because the fundamental color charge of the quark is time dependent in the classical Yang-Mills theory. This is unlike the QED potential which reduces to Coulomb potential at long distance.
\end{abstract}
\maketitle
\pagestyle{plain}

\pagenumbering{arabic}

\section{Introduction}

Among the four fundamental forces of the nature the strong force is responsible for the confinement of quarks (antiquarks) inside the hadron. The color potential $A_\nu^b(t,r)$ produced by the color charge of the quark provides the strong force responsible for the confinement of quarks inside the hadron where $\nu=0,1,2,3$ is the Lorentz index and $b=1,...,8$ is the color index. Hence it is necessary to find the exact form of the color potential $A_\nu^b(t,r)$ produced by the color charge of the quark.

Note that the exact form of the Coulomb electrical potential $V(r)=A_0(r)=\frac{e}{r}$ produced by the electric charge $e$ of the electron at rest is known in the classical Maxwell theory. However, the exact form of the color potential (the Yang-Mills potential) $A_0^b(t,r)$ produced by the color charge of the quark at rest is not known in the classical Yang-Mills theory even if the Yang-Mills theory was discovered in 1954. The determination of the exact form of the color potential $A_\nu^b(t,r)$ produced by the color charge of the quark is a fundamental issue of the strong force of the nature.

It is well known that the quantum electrodynamics (QED) is the quantum field theory of the classical Maxwell theory and the quantum chromodynamics (QCD) is the quantum field theory of the classical Yang-Mills theory. Hence before solving the quantum field theory it is important to solve the corresponding classical theory. This is because while the quantum field theory describes the short distance phenomena of the nature where the quantum loop effects such as vacuum polarizations are important, there are lot of long distance phenomena which can be described by the classical theory because the quantum loop effects such as vacuum polarizations are absent at long distance.

Since the renormalized QED becomes non-perturbative at short distance due to vacuum polarization (quantum loop) effects one finds that the QED potential at short distance differs from the Coulomb form. On the other hand since the vacuum polarization (quantum loop) effects are absent at long distance the QED potential becomes Coulomb potential at long distance. Since the form of the Coulomb potential can be derived from the classical Maxwell theory one finds that the QED potential at long distance is the same potential that is derived in the classical Maxwell theory.

Since Yang-Mills theory was discovered by making analogy with the Maxwell theory by extending U(1) gauge group to SU(3) gauge group \cite{yg,ng1,ng2} one finds that the  form of the QCD potential energy at the long distance and the form of the potential energy in the classical Yang-Mills theory are the same. Since the confinement is a long distance phenomena, the exact form of the Yang-Mills potential (the color potential) $A_\nu^b(x)$ produced by the time dependent color charge $q^a(t)$ of the quark \cite{ng2} in the classical Yang-Mills theory can provide an insight to the question why quarks are confined inside the hadron. Hence it is necessary to find the exact form of the color potential produced by the quark in the classical Yang-Mills theory. The formula for the Yang-Mills potential (the color potential) $A_\nu^b(x)$ produced by the time dependent fundamental color charge $q^a(t)$ of the quark in the classical Yang-Mills theory is derive in \cite{ng1}.

The Coulomb plus linear form of the potential
\bea
V(r)=-\frac{A}{r}+B~r
\label{cpl}
\eea
between color singlet static heavy quark-antiquark pair separated by a distance $r$ is widely used in the literature to study the heavy quarkonium phenomenology where $A$ and $B$ are $r$ independent constants. In case of heavy quarkonium the lattice QCD predicts the Coulomb plus linear form of the potential between the static color singlet heavy quark-antiquark pair \cite{lq}. Similarly many phenomenological models have also used the Coulomb plus linear form of the the potential to study heavy quarkonium \cite{ph}.

However, in this paper we show that the Coulomb potential of the form $\frac{1}{r}$ is not a part of the QCD potential. This is because the form $\frac{g^2}{r}$ is for abelian theory (not QCD) and the form $\frac{g^2(\mu)}{r}$ in QCD at short distance is not of the Coulomb form $\frac{1}{r}$ because  $g(\mu)$ depends on the mass/length scale $\mu$. Similarly at long distance the QCD potential corresponds to the potential in the classical Yang-Mills theory which does not have the Coulomb form $\frac{1}{r}$ because the fundamental color charge $q^a(t)$ of the quark is time dependent \cite{ng2} in the classical Yang-Mills theory. This is unlike the QED potential which reduces to Coulomb potential at long distance. In addition to this the potential energy in eq. (\ref{cpl}) is independent of time whereas the gauge invariant color singlet time dependent potential energy between static quarks in the classical Yang-Mills theory depends on time even if the quarks are at rest \cite{ng3}.

The paper is organized as follows. In section II we show that the path integral formulation of photon field produces coulomb potential energy between static electron-positron pair separated by a distance $r$. In section III we show that the path integral formulation of gluon field does not produce Coulomb potential energy between static quark-antiquark pair separated by distance $r$. In section IV we show that Coulomb potential is not a part of the QCD potential. Section V contains conclusions.

\section{ Path Integral Formulation of Photon Field Produces Coulomb Potential Energy Between Static Electron-Positron }

The effective lagrangian density ${\cal L}(x)$ obtained by using the path integration of the photon field $Q^\nu(x)$ in the presence of external electric current density $j^\nu(x)$ is given by \cite{ng4}
\bea
i \int d^4x {\cal L}(x)={\rm ln} [\frac{\int [dQ] e^{i\int d^4x [-\frac{1}{4}F^2_{\lambda \nu}[Q] -\frac{1}{2\alpha}[\partial_\lambda Q^\lambda(x)]^2 + j_\nu(x) Q^\nu(x)]}}{ \int [dQ] e^{i\int d^4x [-\frac{1}{4}F^2_{\lambda \nu}[Q] -\frac{1}{2\alpha}[\partial_\lambda Q^\lambda(x)]^2]}}]
\label{efla}
\eea
where $\alpha$ is the gauge fixing parameter and
\bea
F_{\lambda \nu}[Q]=\partial_\lambda Q_\nu(x) -\partial_\nu Q_\lambda(x).
\label{fq}
\eea
Using the continuity equation
\bea
\partial_\lambda j^\lambda(x) =0
\label{cne}
\eea
we find from eq. (\ref{efla}) that the effective potential energy $V(r)$ for static charge configuration is given by
\bea
2\int d^4x V(r) = \int d^4x j_\lambda(x) \frac{1}{\partial_\nu \partial^\nu} j^\lambda(x).
\label{eflb}
\eea
For a static electron of charge $e$ at the origin and a static positron at ${\vec r}_1$ we have
\bea
j_\lambda(x) = e \delta_{\lambda 0} \delta^{(3)}({\vec r})-e \delta_{\lambda 0} \delta^{(3)}({\vec r}-{\vec r}_1).
\label{eflc}
\eea
Using eq. (\ref{eflc}) in (\ref{eflb}) and by neglecting the infinite self energies we find that the potential energy $V(r)$ between static electron-positron pair separated by a distance $r$ is given by
\bea
V(r)=-\frac{e^2}{r}.
\label{efld}
\eea
Hence we find that the quantum field theory derivation using the path integral formulation gives the exact form of the Coulomb potential energy between static electron-positron pair which agrees with the corresponding result in the classical Maxwell theory.

\section{ Path Integral Formulation of Gluon Field Does Not Produce Coulomb Potential Energy Between Static Quark-Antiquark }

From eq. (\ref{efla}) we find that the effective lagrangian density ${\cal L}(x)$ obtained by using the path integration of the gluon field $Q_\nu^b(x)$ in the presence of external color current density $j_\nu^b(x)$ is given by
\bea
i \int d^4x {\cal L}(x)={\rm ln} [\frac{\int [dQ] ~{\rm Det}(\frac{\delta \partial^\lambda Q_\lambda^d}{\delta \omega^b})~e^{i\int d^4x [-\frac{1}{4}F^{b^2}_{\lambda \nu}[Q] -\frac{1}{2\alpha}[\partial^\lambda Q_\lambda^b(x)]^2 + j_\nu^b(x) Q^{\nu b}(x)]}}{ \int [dQ]~{\rm Det}(\frac{\delta \partial^\lambda Q_\lambda^d}{\delta \omega^b})~ e^{i\int d^4x [-\frac{1}{4}F^{b^2}_{\lambda \nu}[Q] -\frac{1}{2\alpha}[\partial^\lambda Q_\lambda^b(x)]^2]}}]
\label{qefla}
\eea
where ${\rm Det}(\frac{\delta \partial^\lambda Q_\lambda^d}{\delta \omega^b})$ is the ghost determinant and
\bea
F_{\lambda \nu}^b[Q]=\partial_\lambda Q^b_\nu(x) -\partial_\nu Q^b_\lambda(x)+gf^{bhd}Q_\lambda^h(x)Q_\nu^d(x)
\label{qfq}
\eea
is the non-abelian gluon field tensor. Note that we do not introduce the additional ghost fields but instead work directly with the ghost determinant ${\rm Det}(\frac{\delta \partial^\lambda Q_\lambda^d}{\delta \omega^b})$ in eq. (\ref{qefla}) in this paper.

Due to the presence of the term $gf^{bhd}Q_\lambda^h(x)Q_\nu^d(x)$ in eq. (\ref{qfq}) we find that, unlike eq. (\ref{eflb}) for the abelian case, there is no exact analytical solution of eq. (\ref{qefla}) in the non-abelian case. From eqs. (\ref{efla}) and (\ref{efld}) we find that the only way the eq. (\ref{qefla}) can produce the Coulomb potential energy
\bea
V(r)=-\frac{g^2}{r}
\label{qfld}
\eea
is when
\bea
f^{abc}=0,~~~~~~~~~{\rm for~all}~~~~~~~~~a,b,c=1,...,8
\label{fbc}
\eea
which is equivalent to the color current density [see eq. (\ref{eflc})]
\bea
j_\lambda^b(x) = g{\hat q}^b \delta_{\lambda 0} \delta^{(3)}({\vec r})-g{\hat q}^b \delta_{\lambda 0} \delta^{(3)}({\vec r}-{\vec r}_1)
\label{qflc}
\eea
where
\bea
q^a=g{\hat q}^b
\label{cch}
\eea
is the constant color charge and
\bea
{\hat q}^b{\hat q}^b=1.
\label{nqv}
\eea
Note that in \cite{fsh} the color charge density
\bea
j_0^b(x) = gT^b\delta^{(3)}({\vec r})-gT^b \delta_{\lambda 0} \delta^{(3)}({\vec r}-{\vec r}_1)
\label{fsl}
\eea
was taken which can not be correct because $j_0^b(x)$ is a vector whereas the generator of SU(3) $T^b$ is a matrix with components $T^b_{ij}$. As found out in \cite{ng1,ng2} the form in eq. (\ref{fsl}) can not be a correct form in classical Yang-Mills theory because $j_0^b(x)$ in classical Yang-Mills theory contains infinite powers of $g$. However, the form (\ref{qflc}) is correct in abelian theory \cite{ng1,ng2}.

Hence we find that the form of the Coulomb potential energy in eq. (\ref{qfld}) is for abelian theory (not for QCD). 

\section{ Coulomb Potential Is Not a Part of The QCD Potential }

The origin of the Coulomb potential in QCD is often interpreted as tree level single gluon exchange between static heavy quark-antiquark pair in pQCD at short distance \cite{pqcd}, similar to single photon exchange between static electron-positron pair in pQED at long distance. However, in these tree level single gluon exchange diagram one can not use constant coupling $g$ because for the QCD boils down to a tree level diagram, the QCD coupling constant has to be very weak which can only happen when the asymptotic freedom occurs at very short distance, meaning $g(Q^2)$ is a function of the momentum transfer scale $Q^2$. The tree level single gluon exchange pQCD potential energy between static quark-antiquark separated by a distance $r$ is defined by
\bea
V(r)=-C_F \int \frac{d^3Q}{(2\pi)^3} e^{i {\vec Q}\cdot {\vec r}}~ \frac{\alpha_s(Q^2)}{{\vec Q}^2}
\label{qqp}
\eea
which does not have the Coulomb form $\frac{1}{r}$ because $\alpha_s(Q^2)$ depends on the momentum transfer scale $Q^2$.

Using the running of the QCD coupling one can write eq. (\ref{qqp}) as \cite{qcds}
\bea
V(r)\sim \frac{g^2(\mu)}{r}
\label{qqmu}
\eea
which does not have the Coulomb form $\frac{1}{r}$ because $g(\mu)$ depends on the mass/length scale $\mu$.

At long distance the QCD potential energy is same as the potential energy in the classical Yang-Mills theory \cite{ng3,nka}. For the quark at rest in the limit $g\rightarrow 0$ the Yang-Mills potential (color potential) reduces to \cite{ng1}
\bea
A_0^b(t,r)=\frac{q^b(t-r)}{r},~~~~~~~~~~~~~~~g\rightarrow 0
\label{abl}
\eea
which looks like Coulomb-like potential but is not the Coulomb potential of form $\frac{1}{r}$ because the color charge $q^b(t-r)$ of the quark in eq. (\ref{abl}) depends on $r$. Note that color field also plays an important role in the study of the the quark-gluon plasma \cite{ng5,ng6,ng7,ng8}.

Hence we find that at the long distance the QCD potential corresponds to the potential in the classical Yang-Mills theory which does not have the Coulomb form $\frac{1}{r}$ because the fundamental color charge of the quark is time dependent in the classical Yang-Mills theory

It is useful to mention here that although the static potential $V(r)$ in QED can be defined by the vacuum expectation of Wilson loop \cite{kw,pqcd}
\bea
V(r) = -{\rm lim}_{~T \rightarrow \infty} \frac{1}{T} {\rm ln}<e^{-ie \oint_\Gamma dx^\nu Q_\nu(x)}>
\label{vqed}
\eea
in the Euclidean time formalism but the static potential $V(r)$ in QCD can not be defined by the vacuum expectation of Wilson loop \cite{kw,pqcd}
\bea
V(r) = -{\rm lim}_{~T \rightarrow \infty} \frac{1}{T} {\rm ln}<{\rm tr}~{\cal P}~e^{ig \oint_\Gamma dx^\nu T^bQ_\nu^b(x)}>
\label{vqcd}
\eea
in the Euclidean time formalism because the potential energy $V(t,r)$ between the static quarks is time dependent in the classical Yang-Mills theory even if the quarks are at rest \cite{ng3}.

\section{Conclusions}
The Coulomb plus linear potential is widely used in QCD. However, in this paper we have shown that the Coulomb potential of the form $\frac{1}{r}$ is not a part of the QCD potential. This is because the form $\frac{g^2}{r}$ is for abelian theory (not QCD) and the form $\frac{g^2(\mu)}{r}$ in QCD at short distance is not of the Coulomb form $\frac{1}{r}$ because  $g(\mu)$ depends on the mass/length scale $\mu$. Similarly at long distance the QCD potential corresponds to the potential in the classical Yang-Mills theory which does not have the Coulomb form $\frac{1}{r}$ because the fundamental color charge of the quark is time dependent in the classical Yang-Mills theory. This is unlike the QED potential which reduces to Coulomb potential at long distance.

\end{document}